\documentclass[11pt,a4paper]{article}
\usepackage[utf8]{inputenc}
\usepackage[T1]{fontenc}
\usepackage[english]{babel}
\usepackage{amsmath}
\usepackage{amsfonts}
\usepackage{amssymb}
\usepackage{graphicx}
\usepackage{hyperref}
\usepackage{xcolor}
\usepackage{csquotes}

\usepackage{geometry}
\usepackage[
    backend=biber,
    style=numeric,
    sorting=none,
    giveninits=true
]{biblatex}

\title{Regularity in Cryptocurrency Market Cycles: A Halving-to-Halving Analysis Using Data from Google Trends and Wikipedia}
\author{
	Marek Zatwarnicki \\ \small Independent Researcher \\ \small \href{mailto:marek.zatwarnicki@gmail.com}{marek.zatwarnicki@gmail.com} \and Piotr Stolarski \\ \small Poznań University of Economics and Business \\ \small \href{mailto:piotr.stolarski@ue.poznan.pl}{piotr.stolarski@ue.poznan.pl} \and Włodzimierz Lewoniewski \\ \small Poznań University of Economics and Business \\ \small \href{mailto:wlodzimierz.lewoniewski@ue.poznan.pl}{wlodzimierz.lewoniewski@ue.poznan.pl}
}
\date{}

\begin{document}

\maketitle

\begin{abstract}
	Bitcoin halvings provide a natural temporal structure for studying long-term dynamics of cryptocurrency markets. This paper presents a methodological framework for analyzing the regularity of Bitcoin and aggregate cryptocurrency market cycles by normalizing historical periods between consecutive halving events and projecting them onto a common relative-time scale expressed as the percentage of cycle completion. 
	
	The proposed framework integrates market variables, namely Bitcoin price and aggregate cryptocurrency market capitalization, with measures of public attention derived from Google Trends search volume and Wikipedia pageviews. To obtain long-horizon daily attention indices, Google Trends data are reconstructed from overlapping 90-day windows linked by median scale ratios, following established approaches to combining high-frequency Google Trends samples.
	
	Rather than attempting to forecast future prices, the study aims to compare historical cycles, identify recurring temporal patterns, and evaluate the relationship between market dynamics and retail investor interest across different cycle-relative stages. Exploratory results point to a possible recurring structure. The maxima of the aggregate market capitalization in the two most recent completed cycles occurred at 39.0\% and 37.9\% of cycle completion, and the highest value of the ongoing cycle observed to date at 38.1\% (534--547 days after the respective halvings), although the corresponding multiple of the halving-day value fell from 66 to 1.8. Public attention shows a similar compression of peak multiples, but not a similar timing: attention peaked together with the market in the 2016--2020 cycle, but between 172 and 329 days before the market maximum in the two subsequent cycles. Weekly correlations provide no evidence that attention consistently leads market movements. The paper also discusses current data limitations and outlines future research directions, including formal lead-lag analyses and expanded multi-asset investigations.
\end{abstract}

\vspace{0.5cm}

\section{Introduction}
Cryptocurrency markets are characterized by extreme volatility, rapid structural changes, and recurring boom-and-bust cycles \cite{cheah2015speculative}. Previous research has examined the determinants of Bitcoin price formation and explored various approaches to cryptocurrency valuation and modelling \cite{ciaian2016economics,fantazzini2016everything}. More recent empirical studies have identified recurring price patterns around Bitcoin halving events, suggesting a possible relationship between the halving schedule and longer-term market cycles \cite{fabus2024halving,mbakob2024bubbles}. However, the extent to which these patterns represent stable temporal regularities remains uncertain. Because halvings reduce the rate of new Bitcoin issuance, they provide a natural temporal reference point for comparing successive market cycles.

The persistence of halving-related market cycles remains debated. Recent studies have examined the effects of halving events alongside regulatory, behavioral, and broader market factors, while causal evidence concerning the magnitude and stability of halving effects remains mixed \cite{subawa2025bitcoin, virtonen2025estimating}. While some researchers argue that market efficiency should eliminate predictable halving effects, historical observations indicate that periods of bullish and bearish activity often appear to cluster around successive halving epochs. At the same time, the role of retail investors has become increasingly important as cryptocurrency adoption expanded beyond technically oriented early adopters into mainstream financial markets, and retail attention has been shown to help explain Bitcoin trading activity \cite{aalborg2019what}.

In parallel, the academic literature on blockchain and crypto-assets has expanded and diversified, with recent bibliometric evidence documenting a shift from foundational cryptocurrency topics toward more specialized areas, including decentralized finance, smart contracts, and non-fungible tokens \cite{stolarski2025trends}.

Quantifying retail attention remains challenging, as investor sentiment cannot be measured directly at a global scale. Consequently, proxy metrics such as online search activity and information-seeking behavior have become widely used indicators of public interest \cite{kristoufek2013bitcoin, yelowitz2015characteristics}. Previous studies have documented relationships between online search activity, social media signals and cryptocurrency market dynamics, including price movements, returns, volatility, and trading volume \cite{garcia2014digital, urquhart2018what, nasir2019forecasting, dastgir2019causal}.

This research proposes a unified framework for studying cryptocurrency market cycles within a halving-relative temporal coordinate system. Rather than analyzing market variables in absolute calendar time, all observations are projected onto a normalized timeline expressed as the percentage of completion between two consecutive Bitcoin halvings. This approach enables direct comparison of historical cycles that differ in duration and market maturity. The primary objective of this study is to investigate whether measures of retail attention follow recurring trajectories across market cycles and whether these trajectories exhibit consistent relationships with market capitalization dynamics. To achieve this goal, we combine daily market data for Bitcoin and for the aggregate cryptocurrency market with public attention indicators derived from Google Trends and Wikipedia pageviews. To support open science and reproducibility, the raw and processed datasets are made publicly available in the project's GitHub repository\footnote{\url{https://github.com/piotrstolarski/CryptoMarketCycleAnalysis-research}}.

The main contributions of this work are threefold:
\begin{enumerate}
	 \item A halving-relative framework, combining time and value normalization, that enables direct comparison of cryptocurrency market cycles of different length and scale.

	 \item Exploratory evidence that the market maxima of the two most recent completed cycles, and the highest value of the ongoing cycle to date, occurred at a similar point of the cycle, while their amplitude declined by more than an order of magnitude, together with an assessment of how robust this observation is to the choice of time coordinate.

	 \item An open, daily dataset that aligns market data with long-horizon public attention indices (Google Trends, reconstructed from overlapping windows, and Wikipedia pageviews aggregated over redirects) in a common cycle-relative coordinate system, together with a first comparison of attention and market dynamics across cycles.
\end{enumerate}

The resulting dataset and methodology establish the foundation for a systematic examination of cyclical behavior in cryptocurrency markets and the evolving relationship between public attention and market valuation.

The remainder of this paper is structured as follows. Section 2 details the methodology, including data acquisition, temporal normalization, and the specific algorithms used for Google Trends and Wikipedia data. Section 3 presents the preliminary empirical results regarding peak localization, peak amplitude and drawdowns, and compares the dynamics of public attention with those of the market. Section 4 discusses limitations and directions for future work. Finally, Section 5 concludes the paper.

\section{Methodology}
The proposed methodology consists of four principal stages: (i) acquisition and preprocessing of market and attention data, (ii) reconstruction and normalization of long-term public-interest signals, (iii) transformation of all observations into a halving-based relative-time framework, and (iv) comparative analysis across market cycles.

Market variables comprise the daily closing price of Bitcoin and the aggregate capitalization of the cryptocurrency market. Public attention was measured using Google Trends search activity and Wikipedia pageview statistics. To ensure comparability across multiple market cycles spanning more than a decade, all datasets were synchronized on a daily resolution and normalized where necessary.

\subsection{Market Data}
Daily closing values were obtained from TradingView for the Bitcoin price index (\texttt{INDEX:\allowbreak BTCUSD}, available from 5 October 2009) and for the total cryptocurrency market capitalization (\texttt{CRYPTOCAP:\allowbreak TOTAL}, available from 1 March 2014). The data used in this paper end on 21 July 2026.

Because the first analyzed cycle begins before aggregate market capitalization data are available, the total capitalization before 1 March 2014 was approximated as the product of the Bitcoin price and the estimated circulating supply, with the supply interpolated linearly between the known issuance levels at the genesis block and at each halving. The proxy series was rescaled by a single factor so that it matches the first observed \texttt{CRYPTOCAP:\allowbreak TOTAL} value. The resulting factor, 1.013, is close to unity, which is consistent with Bitcoin accounting for nearly the entire cryptocurrency market at that time. Strictly speaking, however, the series before March 2014 is a proxy for the market capitalization of Bitcoin rather than an observed capitalization of the whole market. Results for the first cycle, in particular its amplitude, are therefore less comparable with those for later cycles.

\subsection{Halving-Based Temporal Normalization}
The analysis covers four cycles delimited by the Bitcoin halvings of 28 November 2012, 9 July 2016, 11 May 2020 and 20 April 2024. For each observation date $t$, a relative cycle coordinate $r(t)$ was computed as
\[ r(t)=\frac{t-t_h^{(n)}}{t_h^{(n+1)}-t_h^{(n)}}, \]

where $t_h^{(n)}$ and $t_h^{(n+1)}$ denote two consecutive Bitcoin halving dates, and $n$ denotes the index of the halving epoch (e.g., $n=1$ for the 2012 halving). The resulting value, expressed as a percentage, represents the degree of cycle completion and allows direct comparison of observations originating from different historical periods.

The three completed cycles lasted 1319, 1402 and 1440 days, respectively. For the ongoing fourth cycle, the next halving date is not yet known; its length is therefore set to the median length of the completed cycles (1402 days). This is an explicit assumption rather than a measurement, and its influence on the results is reported where relevant.

\subsection{Value Normalization}
To compare cycles that differ in nominal scale by several orders of magnitude, each market variable $V$ is expressed as a multiple of its value on the halving day that opens the cycle:
\[ m(t)=\frac{V(t)}{V\big(t_h^{(n)}\big)}, \qquad t_h^{(n)} \le t < t_h^{(n+1)}, \]
so that $m=1$ at the start of every cycle. Multiples are displayed on a logarithmic scale, where equal vertical distances correspond to equal relative changes. Since $m(t)$ depends only on information available at time $t$, the normalization does not introduce look-ahead bias.

\subsection{Google Trends Normalization}
A major challenge in long-term Google Trends analysis is the platform's relative normalization, where each query is independently scaled to the maximum value observed within the requested time interval. Directly combining multiple intervals therefore introduces discontinuities and distorts long-term dynamics.

Several methods have been proposed to construct consistent, long-term Google Trends time series, including the knitting of overlapping high-frequency samples \cite{bleher2022knitting}, the reconciliation of series retrieved at different frequencies \cite{eichenauer2022obtaining}, and calibration against a bank of anchor queries \cite{west2020calibration}. We adopt the overlapping-window idea in a simple form. Daily data are retrieved in windows of $L=90$ days that overlap by $O=30$ days, i.e., with a step of $S=60$ days, covering the period from 1 January 2015 to 6 August 2026 (the last window is shorter). Data were retrieved from Google Trends using the following settings: worldwide geographic coverage, all categories, web search.

For each pair of consecutive windows $P_k$ and $P_{k+1}$, the scale ratio is computed as
\[ r_{k,k+1}=\operatorname{median}_{d\in D_{k,k+1}}\frac{V_k(d)}{V_{k+1}(d)}, \]
where $V_k(d)$ is the value reported for day $d$ in window $P_k$ and $D_{k,k+1}$ is the set of overlap days on which both values are non-zero; days with a zero value in either window are excluded. The ratios are chain-linked forward from the first window, $c_1=1$ and $c_{k+1}=c_k\,r_{k,k+1}$, and each window is rescaled by $c_k/\max_j c_j$, so that the window containing the all-time maximum keeps its original scale and the index reaches 100 at the strongest historical observation. The daily index is the mean of the rescaled values of all windows covering a given day. If an overlap contains no valid day, the scale is carried over unchanged ($r_{k,k+1}=1$). This occurs only for topics with long periods of zero values, which are not analyzed here; for the topics ``Bitcoin'' and ``Cryptocurrency'', every overlap contains 30 valid days.

\subsection{Wikipedia Pageviews Aggregation}
Wikipedia pageview statistics serve as an alternative measure of information-seeking behavior. Compared with search-engine data, Wikipedia traffic provides direct evidence of users actively accessing descriptive information about a cryptocurrency. The potential of Wikipedia and related open knowledge sources in cryptocurrency research has also been explored in previous work. For example, a previous study combined financial data with information from Wikipedia and Wikidata to investigate the success and bankruptcy of cryptocurrency exchanges \cite{stolarski2024exchanges}. While the described study focused on exchange-level risk analysis, the present work examines temporal variations in Wikipedia pageviews as an indicator of public attention.

For each asset, pageview counts were collected from the \texttt{Wikimedia REST API}\footnote{\url{https://wikimedia.org/api/rest_v1/}}. In addition to the canonical article, all English-language redirect pages associated with the asset were identified and aggregated. This procedure reduces fragmentation arising from changes in naming conventions, abbreviations, ticker symbols, and historical article structures.

The resulting daily pageview counts were aggregated into a unified time series and aligned with market and Google Trends datasets for subsequent cycle-based analyses. Pageview data are available from 1 July 2015 to 22 July 2026.

\subsection{Attention Measures Used in the Analysis}
Attention data were collected for Bitcoin, for the general topic of cryptocurrencies, and for a set of major altcoins. The altcoin series are released with the dataset but are not analyzed here, as the market data used in this paper do not include individual altcoins. The analysis focuses on the Google Trends topic and the English Wikipedia article ``Bitcoin''; the topic and article ``Cryptocurrency'' are used as a robustness check. Both sources start before the 2016 halving and therefore cover cycles 2--4; the first cycle is covered only partially and is excluded from the attention analysis.

We addressed two data anomalies before conducting the analysis. First, the Google Trends index for Bitcoin shows an eightfold increase between 30 August and 26 September 2019 that has no counterpart in the Bitcoin price, in Wikipedia pageviews or in the Google Trends topic ``Cryptocurrency''. The increase is present, with identical values, in the raw data of both 90-day windows that cover these days, so it was not introduced by the chain-linking procedure. We regard it as a probable artifact of the retrieved Google Trends sample, although this cannot be verified, and treat the 28 days as missing, replacing them by linear interpolation. This decision affects one result: without it, the Google Trends maximum of cycle 2 would fall on 2 September 2019 (82.0\% of the cycle) instead of December 2017. The results for cycles 3 and 4 and the correlations in Table~\ref{tab:leadlag} change by at most 0.01. Second, Wikipedia pageviews contain short spikes lasting one or two days that affect several articles simultaneously without a corresponding market event (e.g., a four- to sevenfold increase for both ``Bitcoin'' and ``Cryptocurrency'' on 24--25 April 2024) and are likely caused by automated traffic. To make peak detection robust to such spikes, the maximum of each attention series is determined from a centered 7-day median, which removes spikes of up to three days without shifting dates. Even so, the dates of attention maxima are approximate to within several days, and small timing differences should not be interpreted. Attention is normalized in the same way as market variables, i.e., as a multiple of its level on the halving day.

\section{Preliminary Results}
The cycle-relative representation reveals several features that recur across cycles. Throughout this section, the term \emph{maximum} refers to the highest value within a completed cycle; for the ongoing fourth cycle, we use the term \emph{highest value observed to date}, since a higher value may still occur before the next halving.

\subsection{Peak Localization Across Cycles}
\begin{figure}[htbp]
    \centering
    \includegraphics[width=\textwidth]{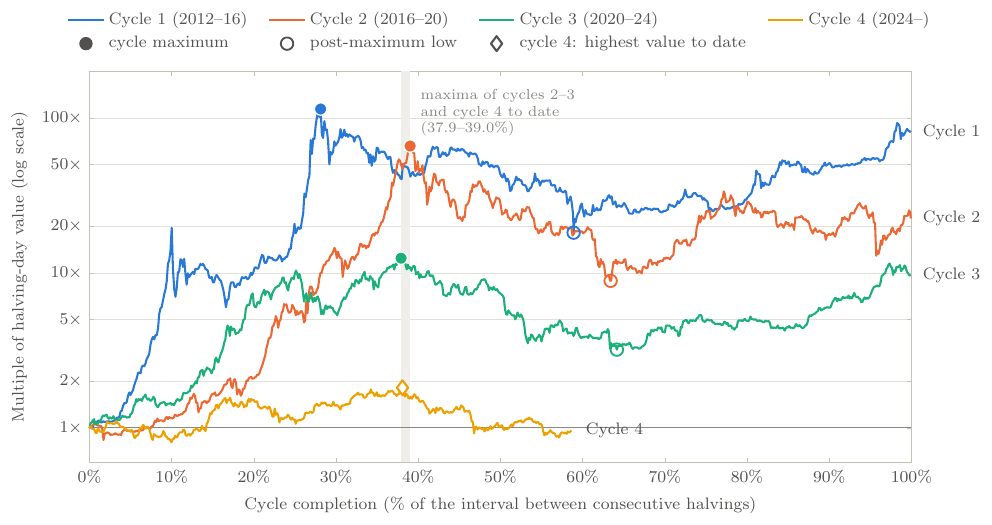}
    \caption{Total cryptocurrency market capitalization in four halving cycles, expressed as a multiple of the halving-day value (logarithmic scale). Filled markers denote the maxima of completed cycles, open circles the lowest value after the maximum, and the open diamond the highest value of the ongoing cycle 4 observed to date. The shaded band spans the maxima of cycles 2 and 3 and the cycle-4 value to date. The length of cycle 4 is assumed to be 1402 days. Source: own calculations based on TradingView data (\texttt{CRYPTOCAP:\allowbreak TOTAL}); before March 2014, a proxy based on the Bitcoin price and estimated supply, so that the first part of cycle 1 reflects Bitcoin only.}
    \label{fig:mcap_cycles}
\end{figure}

One of the most notable observations concerns the temporal location of market maxima relative to Bitcoin halving events. After projecting the aggregate cryptocurrency market capitalization onto the normalized cycle timeline (Figure~\ref{fig:mcap_cycles}), the maxima of the two most recent completed cycles and the highest value of the ongoing cycle to date fall within a narrow interval.

The first completed cycle reached its maximum at 28.1\% of cycle completion (4 December 2013). The second and third completed cycles reached their maxima considerably later, at 39.0\% (7 January 2018) and 37.9\% (8 November 2021), respectively. The highest value of the ongoing fourth cycle observed to date was reached on 6 October 2025, at 38.1\% of the assumed cycle length. Because the length of the fourth cycle is not yet known, this figure is conditional on the 1402-day assumption; a cycle of 1445 days would place the same date at 37.0\%.

Despite large differences in market size, liquidity, institutional participation and the macroeconomic environment, the market maxima may thus occur at a similar point relative to the halving schedule, with a difference of only 1.1 percentage points between cycles 2, 3 and 4, while their amplitude differs by more than an order of magnitude. This similarity rests on two completed cycles and one incomplete cycle, and the first cycle reached its maximum markedly earlier. It should therefore be regarded as an interesting exploratory regularity rather than evidence of a stable cyclical law.

The same pattern is visible for the Bitcoin price alone (Figure~\ref{fig:timing_amplitude}a), with one qualification. In the third cycle, the Bitcoin price exceeded its November 2021 high shortly before the 2024 halving, so the global maximum of that cycle falls at 97.4\% of its duration. The high of November 2021 (7.9 times the halving-day price) is located at 37.9\%, in line with the remaining cycles. For the Bitcoin price, we therefore report the maximum within the first 60\% of each cycle, which coincides with the global maximum in all cycles except the third. The 60\% threshold is an operational choice that excludes the rally immediately preceding the 2024 halving; it is not an independently validated definition of a bull-market top. For the aggregate market capitalization, no threshold is applied and the maximum of the whole cycle is used throughout, so the two measures are not defined identically. Applying the same threshold to the aggregate market capitalization would not change any of its maxima, all of which fall within the first 40\% of the cycle.

\subsection{Diminishing Peak Amplitude}
\begin{figure}[htbp]
    \centering
    \includegraphics[width=\textwidth]{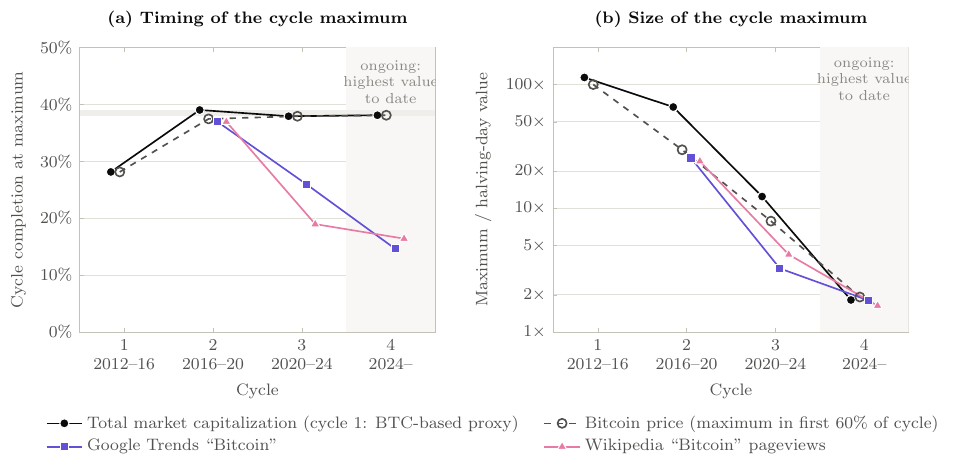}
    \caption{Timing (a) and size (b) of the cycle maximum for market variables and public attention. Timing is expressed as cycle completion at the date of the maximum; size as the maximum divided by the halving-day value. For cycle 4 (shaded column), all values are the highest observed to date. For the Bitcoin price, the maximum within the first 60\% of the cycle is used (an operational threshold); attention maxima are determined from centered 7-day medians. The cycle-1 market capitalization is a Bitcoin-based proxy. Attention data are not available for cycle 1. The shaded band in (a) spans the market capitalization maxima of cycles 2 and 3 and the cycle-4 value to date. Markers are slightly offset horizontally to avoid overlap. Source: own calculations based on TradingView, Google Trends and Wikimedia data.}
    \label{fig:timing_amplitude}
\end{figure}

Although the relative timing of recent market maxima appears similar, their magnitude has decreased substantially over time (Figure~\ref{fig:timing_amplitude}b).

At its maximum, the aggregate market capitalization in the first cycle reached approximately 114 times its halving-day value; as this value rests largely on the Bitcoin-based proxy, it is only partly comparable with later cycles. The second cycle reached 66 times, the third 12.5 times, and the ongoing cycle has so far reached 1.8 times its starting value. The Bitcoin price shows the same sequence at a lower level (100, 30, 7.9 and 1.9 times).

This monotonic decline is consistent with the hypothesis that increasing market maturity and capitalization reduce the magnitude of cyclical expansions. Consequently, any regularity detected in cycle timing should not be interpreted as evidence of comparable future return magnitudes.

\subsection{Post-Peak Drawdown Structure}
A second recurring feature is visible in the drawdown phase following major market tops.

In all three completed cycles, the deepest post-peak decline of the aggregate market capitalization falls in the second half of the normalized timeline, at 58.9\%, 63.4\% and 64.2\% of cycle completion, respectively. The depth of these drawdowns, measured from the cycle peak, was 84\%, 87\% and 74\%. In the ongoing cycle, the lowest value observed after the highest value to date (52\% below it) occurred at 57.1\% of the assumed cycle length, i.e., at the end of June 2026; whether this constitutes the cycle trough cannot yet be determined.

This recurring structure suggests that major periods of market weakness may be consistently aligned in relative-cycle coordinates.

\subsection{Relative Time versus Days Since Halving}
A natural question is whether expressing time as a percentage of cycle completion offers an advantage over simply counting days since the halving. For the maxima of cycles 2 and 3 and the highest value of cycle 4 to date, the two representations perform equally well: these dates fall 547, 546 and 534 days after the respective halvings, a spread of 13 days, compared with a spread of 1.1 percentage points (approximately 15 days) in relative time. The day-based measure has the additional advantage of not depending on the assumed length of the ongoing cycle.

The difference between the two representations increases with the distance from the halving, because differences in cycle length accumulate over time. For the post-peak troughs of the three completed cycles, the spread is 147 days when measured in days since the halving, but 5.3 percentage points (approximately 74 days) in relative time. The relative-time representation thus mainly improves the alignment of the later phases of the cycle, while the location of the market maxima is robust to the choice of time coordinate.

\subsection{Public Attention across Cycles}
\begin{figure}[htbp]
    \centering
    \includegraphics[width=\textwidth]{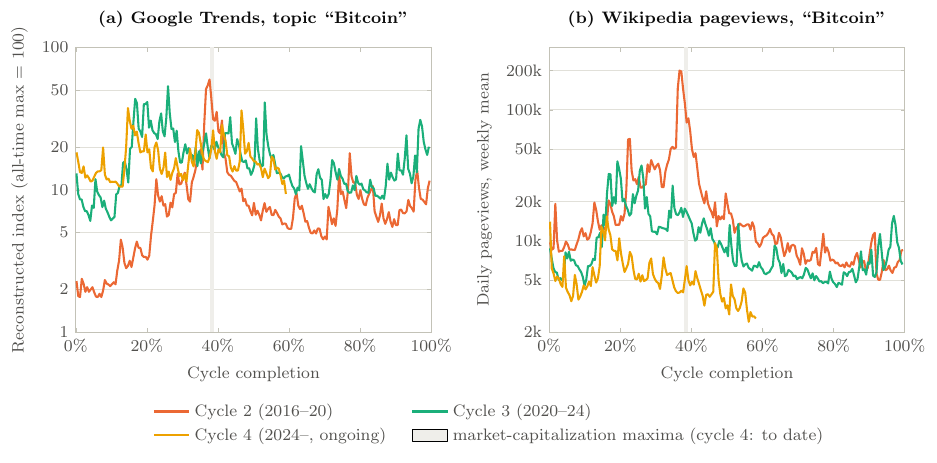}
    \caption{Public attention to Bitcoin in cycles 2--4: reconstructed Google Trends index for the topic ``Bitcoin'' (a) and daily pageviews of the English Wikipedia article ``Bitcoin'', including redirects (b), both as weekly means on a logarithmic scale. The shaded band spans the market capitalization maxima of cycles 2 and 3 and the highest value of cycle 4 to date. The Google Trends values between 30 August and 26 September 2019 are interpolated (see Section~2). Source: own calculations based on Google Trends and Wikimedia data.}
    \label{fig:attention_cycles}
\end{figure}

Figure~\ref{fig:attention_cycles} shows the reconstructed Google Trends index and Wikipedia pageviews for Bitcoin in cycle coordinates. Two observations stand out.

First, the timing of attention maxima differs across cycles, in contrast to the timing of market maxima (Figure~\ref{fig:timing_amplitude}a). In cycle 2, attention and market peaked together: both attention measures reached their maxima on 9 December 2017, at 36.9\% of the cycle and 29 days before the market capitalization maximum. In cycle 3, attention peaked well before the market: Google Trends on 20 May 2021 (26.0\%) and Wikipedia on 8 February 2021 (19.0\%), i.e., 172 and 273 days before the market maximum of November 2021. In cycle 4, the gap widened further. Google Trends peaked on 11 November 2024 (14.6\%) and Wikipedia on 6 December 2024 (16.4\%), two days before Bitcoin first closed above USD 100{,}000. Both dates precede the highest market value to date (6 October 2025) by 329 and 304 days, respectively. In the last two cycles, the market thus continued to rise for several months after public attention had peaked.

These dates are point estimates. They depend on a single Google Trends download, whose sampling variability has not been assessed \cite{medeiros2021proper}, and on the smoothing applied. When maxima are taken from raw daily values instead of 7-day medians, the Wikipedia maximum of cycle 3 moves to 13 May 2021 (179 days before the market maximum), and the Google Trends maximum of cycle 4 moves to a single-day spike on 6 February 2026, during a sharp price decline, 123 days after the highest market value to date. With the exception of this single-day spike, all variants place the attention maxima of cycles 3 and 4 several months ahead of the market maximum.

Second, the compression of peak multiples observed for market variables applies to attention as well (Figure~\ref{fig:timing_amplitude}b). Relative to its halving-day level, the Google Trends index peaked at 25.5, 3.3 and 1.8 times its starting value in cycles 2, 3 and 4, and Wikipedia pageviews at 24.1, 4.2 and 1.6 times, broadly tracking the multiples of the market capitalization. In absolute terms, peak attention declined from cycle to cycle: the 7-day median of daily Wikipedia pageviews peaked at 231{,}000 in cycle 2, 38{,}000 in cycle 3 and 13{,}000 in cycle 4, and the Google Trends index at 62, 48 and 39. Over the same period, the maximum total market capitalization increased from USD 0.69 trillion to USD 2.90 trillion and USD 4.22 trillion (the latter being the highest value to date). This divergence can have several causes, which the present data cannot distinguish. One hypothesis is a growing role of participants whose activity is not reflected in retail-oriented attention measures, such as institutional investors. Other possible explanations include a shift of information seeking from search engines and Wikipedia towards social media, exchange applications and AI assistants, and the greater familiarity of the public with cryptocurrencies, which reduces the need to look up basic information.

To examine short-term dynamics, we computed correlations between weekly log changes of the attention measures and of the total market capitalization at lags of one week (Table~\ref{tab:leadlag}). Weekly values are means over calendar weeks from July 2015 to July 2026 ($n=575$); for independent series, correlations outside $\pm 0.08$ would be significant at the 5\% level. For all four attention measures, correlations with the market change in the same or the preceding week (0.09--0.31) exceed those with the market change in the following week (0.05--0.14). These weekly correlations therefore provide no evidence that attention consistently leads market movements. The pattern is compatible with attention following the market, as reported for Bitcoin by \cite{urquhart2018what}, but simple lagged correlations cannot establish the direction of the relationship. Serial correlation and multiple comparisons are not accounted for.

\begin{table}[htbp]
    \centering
    \caption{Correlation between weekly log changes of attention measures and of the total market capitalization. The column ``market leads'' correlates the change in attention with the market change in the preceding week, ``attention leads'' with the market change in the following week.}
    \label{tab:leadlag}
    \begin{tabular}{lccc}
        \hline
        Attention measure & Market leads & Same week & Attention leads \\
        \hline
        Google Trends, ``Bitcoin'' & 0.18 & 0.09 & 0.07 \\
        Google Trends, ``Cryptocurrency'' & 0.30 & 0.31 & 0.14 \\
        Wikipedia, ``Bitcoin'' & 0.14 & 0.17 & 0.05 \\
        Wikipedia, ``Cryptocurrency'' & 0.10 & 0.20 & 0.06 \\
        \hline
    \end{tabular}
\end{table}

\section{Limitations and Future Work}
The presented framework provides a continuous temporal reference system for cryptocurrency market cycles. Nevertheless, several methodological and data limitations should be noted.

\begin{itemize}
	\item \textbf{Asset Coverage and Entity Disambiguation.} The empirical analysis focuses on Bitcoin and the aggregate market capitalization. While public attention metrics for major altcoins were collected, individual altcoin market data remain to be integrated. Expanding multi-asset coverage also requires resolving search-engine entity disambiguation (e.g., handling multiple topic entities associated with specific tokens such as TRON).

	\item \textbf{Google Trends Sampling Uncertainty.} Reconstructed Google Trends indices are derived from a single retrieved sample, which may exhibit sampling variability \cite{medeiros2021proper}. Future extensions should quantify this uncertainty by aggregating independent extractions.

	\item \textbf{Econometric and Lead-Lag Modeling.} The lead-lag correlations reported in Table~\ref{tab:leadlag} are descriptive and do not account for multiple comparisons. Formal tests that account for serial correlation and volatility clustering (e.g., Granger causality in a vector autoregression), as well as an analysis of the stability of these relationships across cycles, remain to be carried out.

	\item \textbf{Small Sample of Observable Cycles.} Because Bitcoin has completed only three full halving cycles, historical regularities observed across epochs must be interpreted as exploratory patterns rather than definitive cyclical laws.
\end{itemize}

\section{Conclusion}
This work introduces a halving-relative framework for the comparative analysis of cryptocurrency market cycles and retail investor attention. The proposed methodology combines market data, Google Trends search activity, and Wikipedia pageviews within a unified temporal coordinate system based on the percentage of completion between consecutive Bitcoin halving events.

To place retail attention in the same coordinate system, long-term daily Google Trends series were reconstructed from overlapping windows and combined with Wikipedia pageviews aggregated over redirect pages. The resulting dataset provides continuous attention indices covering the last three cycles, including the ongoing one.

Preliminary analysis points to a possible temporal regularity of market maxima. While return magnitudes (peak multipliers) have declined substantially from cycle to cycle, which indicates a clear compression effect as market maturity increases, the timing of recent market maxima has been similar: cycles 2 and 3 reached their maxima at 39.0\% and 37.9\% of cycle completion, and the highest value of cycle 4 to date occurred at 38.1\% (534--547 days after the respective halvings). This similarity is equally visible when time is measured in days since the halving. Because this observation is derived from only two completed cycles and one ongoing cycle---and given that the first cycle peaked markedly earlier---it should be interpreted as an exploratory pattern rather than a definitive law. In addition, the most pronounced post-peak drawdowns occur in the second half of the cycle, and their alignment is clearly improved by the relative-time normalization.

Public attention follows the market in the compression of peak multiples, but not in the timing of peaks. Whereas attention and market peaked together in the 2016--2020 cycle, attention peaked several months before the market in the two subsequent cycles, and its absolute level at the peak declined even as the market capitalization grew. Weekly correlations provide no evidence that attention consistently leads market movements.

These observations should be considered exploratory due to the limited number of completed Bitcoin halving cycles. Future research will focus on statistically quantifying the relationships between public attention and market dynamics through formal lead-lag models, on extending the analysis to individual altcoins, and on multi-asset cycle similarity measurements. Such analyses may help establish whether online attention indicators provide useful information regarding the progression of cryptocurrency market cycles.

\printbibliography[title={References}]

\end{document}